\documentclass[sigconf]{acmart}

\usepackage{subcaption}   
\usepackage{graphicx} 
\usepackage{enumitem} 
\usepackage{pifont} 
\newcommand{\cmark}{\ding{51}} 

\acmConference[EASE 2026]{The 30th International Conference on Evaluation and Assessment in Software Engineering}{9–12 June, 2026}{Glasgow, Scotland, United Kingdom}

\AtBeginDocument{%
  }

\setcopyright{acmlicensed}
\copyrightyear{2018}
\acmYear{2018}
\acmDOI{XXXXXXX.XXXXXXX}
\acmISBN{978-1-4503-XXXX-X/2018/06}

\begin{document}

\title{An Empirical Study of Sustainability in Prompt-driven Test Script Generation Using Small Language Models}

\author{Pragati Kumari}
\affiliation{%
  \institution{University Of Calgary}
  \city{Calgary}
  \country{Canada}}
\email{pragati.kumari@ucalgary.ca}

\author{Novarun Deb}
\affiliation{%
  \institution{University Of Calgary}
  \city{Calgary}
  \country{Canada}}
\email{novarun.deb@ucalgary.ca}

\renewcommand{\shortauthors}{Kumari and Deb}

\begin{abstract}
The increasing use of language models in automated test script generation raises concerns about their environmental impact, yet existing sustainability analyses focus predominantly on large language models. As a result, the energy and carbon characteristics of small language models (SLMs) during prompt-driven unit-test script generation remain largely unexplored. To address this gap, this study empirically examines the environmental and performance trade-offs of SLMs (in the 2B–8B parameter range) using the HumanEval benchmark and adaptive prompt variants (based on the Anthropic template). The analysis uses CodeCarbon to characterize energy consumption carbon emissions and duration under controlled conditions, with unit-test script coverage serving as an initial proxy for generated test quality. Our results show that different SLMs exhibit distinct sustainability profiles: some favor lower energy use and faster execution, while others maintain higher stability or coverage under comparable conditions. Overall, this work provides focused empirical evidence on sustainable SLM-based test script generation, clarifying how prompt structure and model selection jointly shape environmental and performance outcomes.
\end{abstract}




\maketitle

\section{Introduction}
The integration of Large Language Models (LLMs) into automated software testing has significantly advanced test script generation, bridging semantic gaps that traditional symbolic or search-based techniques often struggle to address. However, the deployment of massive foundation models incurs a substantial ``\textit{environmental tax}'' due to high energy consumption and carbon intensity during inference. While most sustainability research in Software Engineering (SE) focuses on these large-scale models, the environmental characteristics of Small Language Models (SLMs)---specifically those in the 2B--8B parameter range---remain under-explored. In resource-constrained environments or privacy-sensitive domains where local execution is preferred, understanding the trade-offs between the energy efficiency of SLMs and their test generation performance is critical for sustainable development.

The primary objective of this research is to empirically evaluate the sustainability-to-performance ratio of SLMs in prompt-driven unit-test script generation. We seek to move beyond traditional accuracy metrics by establishing how prompt engineering and quantization levels influence the physical resource costs of test generation. By utilizing SLMs, we prioritize scientific reproducibility and local control over the computational footprint, addressing the growing need for \textit{Green AI} practices within the software testing lifecycle. Our investigation is guided by the following research questions:

\begin{itemize}[leftmargin=*, labelindent=0pt]
    \item \textbf{RQ-1:} How do different prompt structures and the geographical migration of workloads across regional power grids (with varying carbon intensities) jointly affect the energy consumption (Wh) and the overall carbon footprint (gCO$_2$eq) of SLMs during the generation of unit-test scripts?
    \item \textbf{RQ-2:} Is there a holistic mechanism to rank SLMs for unit-test script generation such that both code coverage and environmental overheads are taken into account?
    \item \textbf{RQ-3:} What are the trade-offs between test coverage and environmental sustainability when employing varying quantization levels (4-bit vs. 8-bit vs. unquantized) across SLM architectures?
    \item \textbf{RQ-4:} Can we give software solution architects a mechanism to prioritize between test coverage and environmental sustainability while choosing an SLM for unit-test script generation? 
    
\end{itemize}

As a direct outcome of this empirical exploration, we define and introduce two novel trade-off metrics: the \textit{Sustainability Velocity Index} (SVI) and the \textit{Green $F_\beta$ score} (GF$_\beta$). SVI measures overall sustainability performance, favoring models that are fast, stable, low-emission, and broadly effective in terms of code coverage. Its multiplicative design prevents high performance in one area from fully compensating for poor performance in another, promoting balanced eco-efficiency. The GF$_\beta$ score is a composite index that balances environmental impact and code coverage, where tuning $\beta$ allows practitioners to prioritize eco-efficiency ($\beta < 1$) or code coverage ($\beta > 1$). 


This empirical analysis revealed that sustainability in unit-test script generation is not merely a function of model size, but is significantly shaped by the interplay between prompt complexity, model quantization, and regional grid carbon intensity. We demonstrate how our proposed SVI and GF$_\beta$ metrics can be used to rank SLMs, identifying configurations that achieve high coverage with minimal environmental impact. The novelty of this work lies in its specific focus on SLMs, its multi-regional carbon analysis, and the introduction of energy-normalized performance metrics. It provides actionable evidence for selecting model-prompt configurations that balance testing efficacy with environmental responsibility in a global context.

The remainder of the paper is structured as follows: Section \ref{sec:related} documents the related works and gaps in the existing literature. Section \ref{sec:methodology} presents the empirical study pipeline, followed by Section \ref{sec:experimental_setup} that documents the experimental setup, results and discussion. Section \ref{sec:threats} discusses the
threats to validity of the proposed study. Finally, Section \ref{sec:Conclusion} concludes the paper.
\vspace{-1ex}
\section{Related Work}\label{sec:related}

Recent work has explored sustainability in prompting and code generation. Studies such as \cite{wu2023greenprompting,cappendijk2025generating} introduced prompt-based techniques to reduce emissions at inference time. These works evaluated structured prompts across various language models (e.g., GPT-3, Code Llama, DeepSeek-Coder) and demonstrated that prompt phrasing can impact energy usage without compromising output quality. In addition, the influence of prompting patterns on sustainability and model efficiency has been examined \cite{oprescu2023prompt}, offering structured perspectives on energy-aware prompt engineering.

Efforts to model and monitor carbon emissions throughout the lifecycle of language models are presented \cite{deng2024llmcarbon,luccioni2023ml}. The corresponding tools and frameworks, including LLMCarbon and ML Bloom, provide lifecycle-aware and real-time tracking of emissions, enhancing transparency and enabling sustainability-informed development decisions.

The environmental impact of automated test script generation has been assessed \cite{li2023energy,sharma2023sustainable}. These studies analyzed energy usage patterns using models such as CodeT5, GraphCodeBERT, and quantized SLMs, highlighting key considerations in emission-aware testing. Work such as \cite{fraser2011evosuite} introduced EvoSuite, an evolutionary testing tool targeting branch coverage, though without a focus on energy consumption.

Two key reviews \cite{lago2014systematic,mourao2018green} classified green software metrics and sustainability practices, emphasizing the absence of integrated sustainability metrics in early phases of the software lifecycle and advocating for tool-supported green engineering methodologies. Further extensions to sustainability-focused quality frameworks have been proposed \cite{kapoor2024green}, incorporating maintainability and energy efficiency into software development processes.

Sustainability in large language models has also been examined from operational, deployment, and educational perspectives. Studies such as \cite{iyer2023cloud,verdecchia2021green,vartziotis2024learn} analyzed real-world usage patterns, prompting strategies, and organizational interventions aimed at improving energy and emissions awareness in cloud-hosted AI and code generation contexts. Broader industry guidance is provided through initiatives such as \cite{gsf2025}, which promote best practices and standards for carbon-aware software development. Foundational analyses of energy and carbon costs in NLP tasks were presented \cite{strubell2019energy,rashkin2021words}, establishing relationships between batch size, model configuration, task type, and resulting carbon intensity.

While prior studies explore sustainability in prompting, LLM inference, and test generation, none specifically examine the environmental characteristics of small language models during automated test script generation. Existing frameworks largely reflect assumptions aligned with large-model behavior, overlooking SLM-specific aspects such as time-aware emissions, prompt sensitivity, and run-to-run stability. This gap underscores the need for an evaluation approach tailored to SLM inference patterns. The empirical study performed in this paper addresses this need by offering a multidimensional sustainability assessment designed specifically for SLM-based unit-test script generation.

\section{The Empirical Study}
\label{sec:methodology}

\begin{figure*}[!ht]
    \centering
    \includegraphics[width=\textwidth]{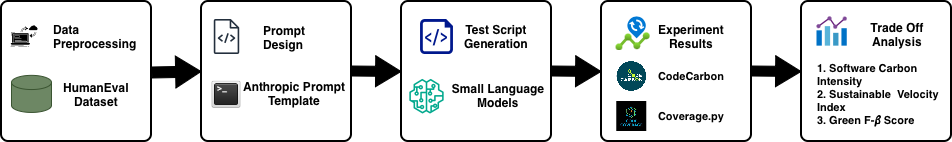}
    \caption{The Empirical Study pipeline}
    \label{fig:workflow}
\end{figure*}

To systematically evaluate the sustainability and performance trade-offs of using compact language models for test automation, we propose an empirical study pipeline illustrated in Figure \ref{fig:workflow}. The workflow is structured into five core phases: (1) \textit{Data Preprocessing}, (2) \textit{Prompt Design}, (3) \textit{Test Script Generation}, (4) \textit{Experimental Results}, and (5) \textit{Sustainability Trade-off Analysis}. 
The following subsections detail these phases.

\subsection{Data Preprocessing}


The foundation of our experimental evaluation is the widely adopted \texttt{HumanEval} benchmark \footnote{https://huggingface.co/datasets/openai/openai\_humaneval}, a standardized dataset comprising 164 Python programming tasks used to measure the coding capabilities of language models of varying parameter sizes. Each entry in the \texttt{HumanEval} dataset is structured as a record containing five key components:

\begin{enumerate}[label=(\roman*)]
  \item \textit{task\_id}: a unique identifier for each problem;
  \item \textit{prompt}: the natural language problem statement and the required Python function signature;
  \item \textit{canonical\_solution}: a reference implementation of the correct function logic;
  \item \textit{test}: an assertion-based validation script containing the pre-defined, human-written unit tests; and
  \item \textit{entry\_point}: the callable name of the function to be tested.
\end{enumerate}

For this work, a crucial preprocessing step was necessary to prepare the data for the test script generation and evaluation process. We merged the \textit{prompt} (defining the task and signature) and the \textit{canonical\_solution} (the target code under test) to create a single runnable Python code file for each of the 164 tasks. This unified file represents the \textit{Software Module Under Test} \footnote{\url{https://github.com/Kumari-Pragati/Sustanability_SLMs_Study}}.
The functions within the \texttt{HumanEval} dataset, paired with their corresponding ground-truth unit tests, serve as the foundation for the inference pipeline. 

\subsection{Prompt Design} \label{subsec:PrmptVar}

To systematically investigate the influence of instructional complexity on both the coverage of the test script and the environmental cost of inference, we designed four progressively constrained prompt templates (AP$_{V_0}$ through AP$_{V_3}$). This structured approach adheres to industry-standard prompt engineering methodologies, specifically leveraging the steering and constraint techniques proposed by Anthropic\footnote{https://www.youtube.com/watch?v=ysPbXH0LpIE}. 
The structural evolution of these variants is summarized in Table~\ref{tab:prompt_variants}. While the baseline $AP_{V_0}$ provides only the raw function signature and docstring, the subsequent variants introduce sophisticated steering mechanisms. $AP_{V_1}$ establishes an explicit \textit{``Expert Tester''} persona to align the model's internal weights; $AP_{V_2}$ introduces structured rules (DOs/DON'Ts) and strictly separates System and User roles for refined output control; and $AP_{V_3}$ provides the most structured prompt by  introducing explicit requirements for test methods, limits on tone/format, and providing an illustrative example block to ground the model's output format. This progression allows us to measure the \textit{environmental tax} of adding instructional tokens relative to the gains in functional test coverage. The complete set of prompt templates used in this study is available in our replication repository\footnote{\url{https://github.com/Kumari-Pragati/Sustanability_SLMs_Study}}.

\begin{table}[!h]
\caption{Comparative analysis of the structural complexity of the prompts derived from the Anthropic template.}
\label{tab:prompt_variants}
\centering
\footnotesize
\setlength{\tabcolsep}{4pt} 
\begin{tabular}{@{} l c c c c @{}}
\toprule
\textbf{Features} & \textbf{AP$_{V_0}$} & \textbf{AP$_{V_1}$} & \textbf{AP$_{V_2}$} & \textbf{AP$_{V_3}$} \\
\midrule
Minimal runnable template            & \cmark & \cmark & \cmark & \cmark \\
Explicit ``expert tester'' persona   &        & \cmark & \cmark & \cmark \\
Structured rules (DOs/DON'Ts)        &        &        & \cmark & \cmark \\
System/user role split               &        &        & \cmark & \cmark \\
Required test methods                &        &        &        & \cmark \\
Tone/format constraints               &        &        &        & \cmark \\
Illustrative example block            &        &        &        & \cmark \\
\bottomrule
\end{tabular}
\end{table}

\subsection{Test Script Generation}

The Test Script Generation phase focuses on the systematic creation of unit-test scripts by the selected SLMs based on the structured prompt variants ($AP_{V_0}$ to $AP_{V_3}$). For this study, we selected five state-of-the-art SLMs (\texttt{Phi-3.5-mini}, \texttt{Qwen2.5-1.5B}, \texttt{deepseek-coder-\\7b}, \texttt{Mistral-7B}, and \texttt{Llama-3-8B}) within the 2B--8B parameter range. These SLMs were instantiated under different quantization schemes to evaluate the trade-off between model compression and environmental sustainability. The model configurations are structured as follows:

\begin{itemize}
    \item \textbf{Unified 8-bit Evaluation:} To establish a consistent baseline for comparison across different architectures, all five SLMs were tested using a unified 8-bit quantization scheme.
    \item \textbf{Variable Quantization Analysis:} To specifically investigate the impact of model compression on sustainability, two models---\texttt{Phi-3.5-mini} and \texttt{Qwen2.5-1.5B}---were subjected to a comparative analysis across three distinct levels: 4-bit, 8-bit, and unquantized.
\end{itemize}

This tiered approach allows us to isolate the effects of architectural differences under a fixed quantization level while simultaneously exploring how varying levels of precision affect the energy-to-coverage ratio within specific models. All generation trials were conducted under a controlled environment to ensure that observed variations in sustainability metrics are attributable to the experimental variables.

\subsection{Experiment Results}
To assess the environmental impact of SLMs in automated unit-test script generation, our study records four \textit{primary metrics} derived from fundamental execution parameters: (i) \textit{Test Coverage} ($Q$); (ii) \textit{Energy Consumption} ($E$); (iii) \textit{Carbon Emissions} ($C$); and (iv) \textit{Duration} ($T$). To ensure empirical rigor, we utilize the \texttt{CodeCarbon} library to track $E$, $C$, and $T$, while the \texttt{Coverage.py} tool is used to quantify the code coverage of the generated unit-test scripts ($Q$). Furthermore, since our experiments are conducted on Google Colab -- which dynamically migrates workloads to data centers globally---we systematically collect the regional data (\texttt{country\_iso\_code}) for each trial. This metadata is critical for our analysis, as it accounts for the varying percentages of renewable energy in different national power grids. Together, these primary data points form the basis for the multi-dimensional trade-off analysis and the calculation of our novel sustainability metrics described in the following section.

\subsection{Trade Off Analysis}

Building on the fundamental execution parameters, we compute derived indices that explicitly quantify the multi-dimensional trade-offs between test coverage and environmental impact. They provide a robust quantitative basis for selecting model-prompt configurations that optimize both functional coverage and ecological responsibility.

\subsubsection{Software Carbon Intensity (SCI)}

The \textit{Software Carbon Intensity (SCI)} metric, adapted from the Green Software Foundation's standard \cite{gsf2025}, represents the environmental cost of a model's computational activity as the amount of \textit{grams of $\text{CO}_2$ equivalent ($\text{gCO}_2\text{e}$)} emitted per functional run. It is directly proportional to the total energy consumed (by CPU, GPU, and RAM) and can be expressed as:
\begin{equation}
\label{eq:sci}
\text{SCI} \propto E \;\implies\; \text{SCI} = kE,\ \text{where } k = \frac{I}{R}.
\end{equation}

Here, the proportionality constant $k$ is defined by the fixed environmental and operational parameters of the experiment.
\begin{itemize}
    \item $E$ (in $\mathrm{kWh}$) denotes the \textit{total energy consumption} recorded during the entire batch of test generation.
    \item $I$ (in $\mathrm{gCO}_2/\mathrm{kWh}$) is the \textit{carbon intensity} of the power grid where the inferences were performed by Google Colab.
    \item $R$ is the \textit{number of functional runs} (i.e., the number of individual Python code modules from the \texttt{HumanEval} dataset processed within that batch).
\end{itemize}

\vspace{2pt}
\noindent
\textit{SCI} measures the carbon emissions per generated test script (one execution cycle for a single code module); lower values indicate greater energy efficiency. It serves as the baseline metric in our sustainability study for emission-aware performance comparison and captures the ``\textit{carbon footprint of productivity}'' in line with ISO~14067 and Green Software Foundation guidance \cite{gsf2025}.

\subsubsection{Sustainable Velocity Index (SVI)}

The \textit{Sustainable Velocity Index (SVI)} combines four critical sustainability aspects\textemdash coverage, software carbon intensity, execution time, and stability\textemdash into a single normalized metric.

\begin{equation}
\label{eq:svi}
\text{SVI} = \frac{Q}{100} \cdot \text{ECO}, 
\qquad
\text{ECO} = \frac{1}{(1+\hat{\text{SCI}})(1+\hat{T})}
\end{equation}

In this formula:
\begin{itemize}
    \item $Q$ is the code coverage achieved (\%).
    \item $\text{ECO}$ represents the eco-efficiency index, combining normalized $\text{SCI}$ ($\hat{\text{SCI}}$) and normalized execution time ($\hat{T}$), where $T$ is the duration in seconds.
\end{itemize}

\textit{SVI} summarizes overall sustainability by combining coverage ($Q$), normalized emissions ($\hat{\text{SCI}}$), and normalized runtime ($\hat{T}$), where higher values indicate better balanced performance. Its multiplicative structure prevents any strength from masking a major weakness. All normalizations are done using Min-Max Normalization.

\subsubsection{Green $\text{F}_{\beta}$ Score ($\text{GF}_{\beta}$)}

The \textit{Green $\text{F}_{\beta}$ Score ($\text{GF}_{\beta}$)} synthesizes eco-efficiency ($\text{ECO}$) and code coverage accuracy ($Q$), balancing environmental impact with performance.

\begin{equation}
\label{eq:gfbeta}
\text{GF}_{\beta} = \frac{(1+\beta^{2})Q \cdot \text{ECO}}{\beta^{2}Q + \text{ECO}} 
\end{equation}

Here, $\beta$ is a weighting parameter that controls the trade-off between code coverage and eco-efficiency. The mean values of $\text{GF}_{\beta}$ are compared in two regimes: $\beta < 1$ (eco-efficiency domain) and $\beta > 1$ (code coverage-oriented domain). Higher $\text{GF}_{\beta}$ when $\beta < 1$ indicates greater eco-efficiency under time and emission constraints, while higher $\text{GF}_{\beta}$ when $\beta > 1$ reflects better code coverage under ecological limitations. This tunability supports sustainability analysis by allowing stakeholders to choose the balance that best matches their environmental and performance priorities.
$\text{GF}_{\beta}$ integrates the dynamics of time, emission and code coverage into a single composite index, allowing a holistic evaluation of how models optimize resource usage while maintaining accuracy of coverage. 


\section{Experiments, Results and Discussion}
\label{sec:experimental_setup}

To ensure full reproducibility and isolate the effects of prompt structure and quantization on sustainability, all experiments were conducted under the unified configuration and sequential pipeline summarized in Table~\ref{tab:experimental_config}.
\begin{table}[h]
\centering
\caption{ Experimental Parameters and Setup}
\label{tab:experimental_config}
\footnotesize
\begin{tabular}{@{}llp{4.5cm}@{}}
\toprule
\textbf{Category} & \textbf{Parameter} & \textbf{Specification / Value} \\ \midrule
\textbf{Compute} & Hardware & NVIDIA T4 GPU (16 GB) via Google Colab \\
 & Software Stack & \texttt{transformers}, \texttt{accelerate}, \texttt{bitsandbytes}, \texttt{tqdm} \\ \midrule
\textbf{Monitoring} & Sustainability & \texttt{CodeCarbon} (v3.0.0) \\
 & Power Grid Intensity & 3-month regional average (Electricity Maps) \\
 & Coverage & \texttt{Coverage.py} (v7.13.4) \\ \midrule
\textbf{Inference} & Parameters & Temp: 0.0; Max Tokens: 1024; \\
 & Batch Size & 5 code files per batch \\
 & \#Batches &  33 batches per run \\
 & \#epochs &  3 per run \\
 & Warm-up &  5 minutes (Fibonacci Sequence) \\
 & Cool-down &  1 minute \\
 & Prompt Variants & $AP_{V_0}, AP_{V_1}, AP_{V_2}, AP_{V_3}$ \\ \midrule
\textbf{Evaluation} & Data Granularity & 20 CSV logs (5 models $\times$ 4 prompt variants) \\
 & $GF_\beta$ Weights & $\beta \in \{0.01, 1.0, 10.0\}$ \\ \bottomrule
\end{tabular}
\end{table}
The pipeline execution consists of six stages: (i) \textit{Warm-up}, a 5-minute Fibonacci-number generation task; (ii) \textit{Batch execution}, generating test scripts across all model-prompt combinations while logging resource consumption; (iii) \textit{Data consolidation}, merging the 20 CSV logs into a unified dataset of emissions (converted to gCO$_2$), energy (kWh), and time (s); (iv) \textit{Coverage measurement}, computing average \% coverage per configuration; (v) \textit{Sustainability Ranking}, applying the formulas for SVI and $GF_\beta$ to the aggregated data; and (vi) \textit{Cool-down}, a 10-minute post-run period. All specific calculation steps and replication scripts are available in our repository\footnote{\url{https://anonymous.4open.science/r/Sustanability_SLMs_Study-8668}}.

\subsection{Results}
\label{sec:results}
Here we present the empirical results of our study, analyzing the sustainability trends and performance trade-offs observed across the evaluated SLM architectures, quantization levels, and prompt variants.

\begin{figure}[b]
    \centering
    \includegraphics[width=\columnwidth]{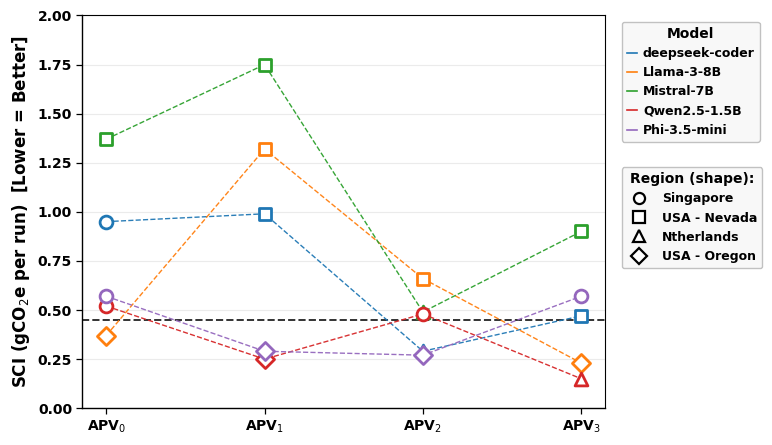}
    \caption{SCI evaluations for different model-prompt configurations w.r.t. the geographic region where the inference was performed by Google Colab.}
    \label{fig:sci}
\end{figure}

\subsubsection{Software Carbon Intensity (SCI)}

Figure~\ref{fig:sci} summarizes SCI for all five models across AP$_{V_0}$--AP$_{V_3}$. When runs occur in the same region (e.g. \texttt{Mistral-7B} in USA--Nevada for AP$_{V_0}$ and AP$_{V_1}$, or \texttt{Phi-3.5-mini} in USA--Oregon for AP$_{V_1}$ and AP$_{V_2}$), SCI differences primarily reflect prompt-model configuration overhead; when regions change, grid carbon intensity becomes the dominant driver. In our setting, SCI values below the 0.45 gCO$_2$e median align with lower-intensity grids (Netherlands: 153 gCO$_2$e/kWh; USA--Oregon: 207 gCO$_2$e/kWh) versus higher-intensity grids (USA--Nevada: 472 gCO$_2$e/kWh; Singapore: 477 gCO$_2$e/kWh).

\vspace{2pt}
\noindent
\colorbox{gray!15}{
\parbox{\dimexpr\linewidth-2\fboxsep}{%
\textbf{Answer to RQ-1:} SCI is jointly shaped by (i) prompt-driven inference overhead (energy use) and (ii) the carbon intensity of the execution region.
When prompt variants run in the same region, SCI changes can be attributed more directly to prompt/model effects. When regions differ, grid intensity can dominate.
}}

\subsubsection{Sustainable Velocity Index (SVI)}

\begin{figure}[h]
    \centering
    \includegraphics[width=\columnwidth]{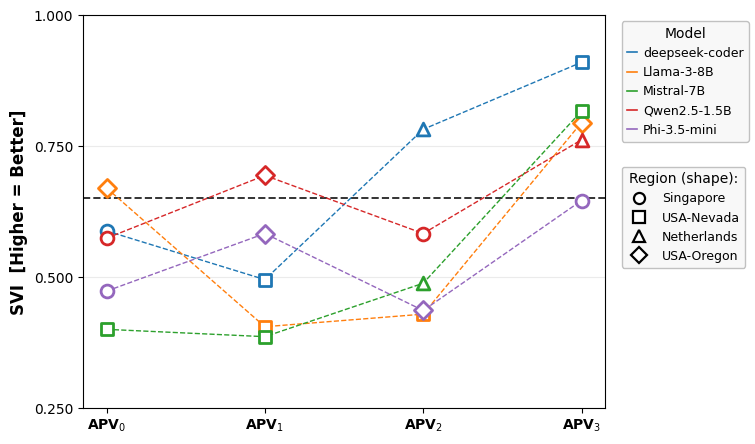}
    \caption{SVI scores for different model-prompt configurations w.r.t. the geographic region where the inference was performed by Google Colab.}
    \label{fig:svi_main}
\end{figure}

Figure~\ref{fig:svi_main} presents SVI trends for all five SLMs (with a unified 8-bit quantization) in the four prompt variants. The median reference line at SVI$=0.65$ provides a useful aggregate split between lower- and higher-sustainability configurations. Most points above the median occur when runs migrate to the Netherlands or USA--Oregon, suggesting that lower-carbon grids generally help lift the SVI, although the separation is less visually pronounced than in the SCI plot (Figure~\ref{fig:sci}). This is because SVI also mixes coverage, stability, and time. Notably, there is overlap across regions: some USA--Nevada runs still exceed the median (e.g., \texttt{Mistral-7B} and \texttt{deepseek-coder-7b} under AP$_{V_3}$), while some Netherlands/USA--Oregon runs remain below the median (e.g., \texttt{Phi-3.5-mini} under AP$_{V_1}$ and AP$_{V_2}$, and \texttt{Mistral-7B} under AP$_{V_2}$).

\vspace{2pt}
\noindent
\colorbox{gray!15}{
\parbox{\dimexpr\linewidth-2\fboxsep}{%
\textbf{Answer to RQ-2:} 
SVI provides a holistic ranking signal by jointly integrating coverage, stability, time, and emissions into a single score. 
While lower-carbon regions tend to dominate above-median outcomes, overlaps show that prompt-model effects can counterbalance reegional power grid intensities, reinforcing the need for multidimensional scoring rather than any single metric.
}}

\begin{figure}[t]
    \centering
    \includegraphics[width=\columnwidth]{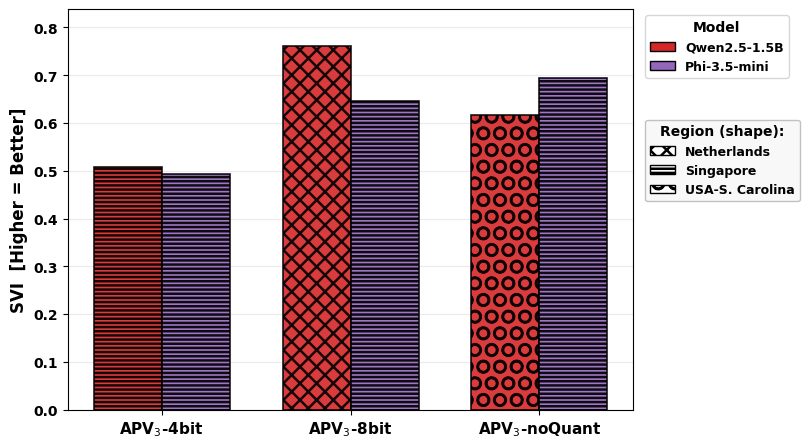}
    \caption{SVI scores for different quantizations of \texttt{Phi-3.5-mini} and \texttt{Qwen2.5-1.5B} w.r.t. the geographic region where the inference was performed by Google Colab.}
    \label{fig:svi_quant}
\end{figure}

\subsubsection{Quantization Sensitivity Analysis (AP$_{V_3}$)}

Figure~\ref{fig:svi_quant} summarizes the impact of numerical precision on the balance between sustainability and coverage. 
For \texttt{Phi-3.5-mini}, all runs were executed in Singapore (constant grid intensity), and the SVI increases from 0.494 (4-bit) to 0.646 (8-bit) to 0.694 (unquantized), indicating a clear improvement with higher precision under a fixed regional context. In contrast, \texttt{Qwen2.5-1.5B} was executed across three regions and exhibits a non-monotonic pattern: 8-bit in the Netherlands achieves the highest SVI (0.762; 153 gCO$_2$e/kWh), while unquantized in USA--South Carolina is lower (0.617; 471 gCO$_2$e/kWh) and 4-bit in Singapore is lowest (0.507; 477 gCO$_2$e/kWh), highlighting how regional grid intensity can dominate the apparent effect of quantization.

\vspace{2pt}
\noindent
\colorbox{gray!15}{
\parbox{\dimexpr\linewidth-2\fboxsep}{%
\textbf{Answer to RQ-3:} 
Quantization introduces a model- and context-dependent trade-off: in a fixed region, higher precision improved SVI, suggesting coverage gains can outweigh any compute savings from lower bits. 
However, when execution region changes, grid intensity can dominate the outcome, so quantization effects must be interpreted alongside location-dependent emissions.
}}


\subsubsection{Green F-$\beta$ Score (GF$_\beta$)}
Table~\ref{tab:beta_ranking} presents model ranking for $\beta \in \{0.01, 1.0, 10\}$ and prompt variants AP$_{V_0}$ and AP$_{V_3}$ (lower rank = better). The choice of $\beta$ (sustainability- vs.~coverage-emphasis) can materially change the relative ordering of SLMs under GF$_\beta$, i.e., models that rank best under sustainability-focused settings do not necessarily remain best when coverage is prioritized. Moreover, even for the same $\beta$, rankings can shift between AP$_{V_0}$ and AP$_{V_3}$ because prompt variants and model--prompt configurations are not consistently executed in identical regions; this introduces location-dependent emissions differences that can alter the composite ranking.

\vspace{2pt}
\noindent
\colorbox{gray!15}{
\parbox{\dimexpr\linewidth-2\fboxsep}{%
\textbf{Answer to RQ-4:} 
To support solution architects in prioritizing coverage versus sustainability, GF$_\beta$ provides an explicit ``\textit{knob}'' ($\beta$) that re-weights the ranking toward eco-efficiency (lower $\beta$) or code coverage retention (higher $\beta$). The chosen $\beta$ and the deployment context (prompt and region) can reshuffle model orderings, so model selection should be performed with the intended priority setting and expected execution location in mind.
}}

\begin{table}[t]
\centering
\footnotesize
\captionsetup{font=footnotesize}
\setlength{\tabcolsep}{3pt}
\renewcommand{\arraystretch}{1.05}
\caption{Model ranking across $\beta$ for AP$_{V_0}$ and AP$_{V_3}$ (lower = better).}
\begin{tabular}{lcccccc}
\toprule
Model & \multicolumn{2}{c}{$\beta=0.01$}
      & \multicolumn{2}{c}{$\beta=1$}
      & \multicolumn{2}{c}{$\beta=10$} \\
\cmidrule(lr){2-3}\cmidrule(lr){4-5}\cmidrule(lr){6-7}
 & AP$_{V_0}$ & AP$_{V_3}$
 & AP$_{V_0}$ & AP$_{V_3}$
 & AP$_{V_0}$ & AP$_{V_3}$ \\
\midrule
Phi-3.5-mini   & 1 & 1 & 1 & 3 & 1 & 5 \\
Qwen2.5-1.5B & 2 & 3 & 4 & 5 & 5 & 4 \\
Llama-3-8B  & 4 & 3 & 3 & 2 & 3 & 2 \\
Mistral-7B& 5 & 3 & 5 & 4 & 4 & 3 \\
DeepSeek-coder-7b  & 3 & 2 & 2 & 1 & 2 & 1 \\
\bottomrule
\end{tabular}
\label{tab:beta_ranking}
\end{table}

The results discussed above indicate that the sustainability dynamics differs between the sensitivity of emissions and the efficiency of test-script generation in terms of code coverage. In general, no single model dominates all sustainability dimensions; optimal selection depends on whether emission minimization or coverage efficiency is prioritized in region-sensitive deployment settings.





\section{Threats to Validity}
\label{sec:threats}

This study presents several potential threats to validity.

\textit{Internal Validity:} Although the experimental pipeline was kept constant, uncontrolled factors such as system background processes, variability in model loading, or cache effects may have influenced the energy measurements and execution time, potentially introducing bias in reported outcomes.

\textit{External Validity:} The generalizability of the results is limited due to the reliance on a single benchmark (HumanEval), fixed hardware configuration, and specific assumptions of the carbon intensity of electricity. Performance and emission outcomes may vary across other datasets, programming tasks, geographic regions (owing to electricity grid differences), or hardware architectures such as TPU, AMD GPU, or edge devices.

\textit{Construct Validity:} Emission values were derived using estimated carbon intensity from CodeCarbon rather than real-time smart-meter measurements. Therefore, the reported results serve as approximations rather than absolute emission values. Similarly, test coverage and accuracy may not fully capture the broader notions of quality, usability, or robustness of AI-generated test scripts.

\textit{Statistical Conclusion Validity:} Experiments were carried out with a limited number of trials per model–prompt pair, which may affect the reliability of observed trends in emission variations and time–efficiency relationships. Minor fluctuations might not reach statistical significance, and potential correlations should be interpreted with caution.




\section{Conclusion}\label{sec:Conclusion}
This study presented a comprehensive sustainability evaluation of small language models (SLMs) for prompt-driven test script generation using the HumanEval dataset and progressive prompt variants (AP$_{V_0}$--AP$_{V_3}$). The evaluation characterizes inference-time sustainability in a multidimensional way by jointly considering energy and carbon costs alongside execution time, run-to-run stability, and code coverage trade-offs, enabling practical comparison of models under consistent experimental conditions.

Experimental results revealed clear differences across the studied SLMs. These findings highlight that sustainability must be evaluated across multiple complementary dimensions rather than through a single metric. No model consistently outperforms others in every aspect; instead, each exhibits distinct strengths. Consequently, model selection should be guided by the specific objective of deployment—whether the emphasis is on reducing emissions, ensuring stable execution behavior, or maintaining stronger coverage–performance balance.

Future work will further investigate prompt structure impacts on inference sustainability, extend the evaluation to additional model families and benchmark and examine quantization and hardware-level variations to better understand sustainable deployment trade-offs.

\bibliographystyle{ACM-Reference-Format}
\bibliography{sample-base}










\end{document}